\documentstyle[preprint,aps]{revtex}

\tightenlines

\begin{document}

\draft

\title{
Mirror matter admixtures in $K_{\rm S}\to\gamma\gamma$
}

\author{
G.~S\'anchez-Col\'on
}
\address{
Departamento de F\'{\i}sica Aplicada.\\
Centro de Investigaci\'on y de Estudios Avanzados del IPN. Unidad Merida.\\
A.P. 73, Cordemex. M\'erida, Yucat\'an, 97310. MEXICO.
}
\author{
A.~Garc\'{\i}a
}
\address{
Departamento de F\'{\i}sica.\\
Centro de Investigaci\'on y de Estudios Avanzados del IPN.\\
A.P. 14-740. M\'exico, D.F., 07000. MEXICO.
}

\date{\today}

\maketitle

\begin{abstract}
The latest measurement of the $K_{\rm S}\to\gamma\gamma$ branching
ratio clearly shows an enhancement over the current theoretical
prediction. As in other $K$ and $B$ meson decays, this invites to consider
the possibility of the contribution of new physics. We study a particular
form of the latter, which may be referred to as manifest mirror
symmetry. The experimental data are described using previously determined
values for the mixing angles of the admixtures of mirror matter in
ordinary hadrons and by assuming that for
$\pi^0,\eta,\eta'\to\gamma\gamma$, the mirror decay amplitudes have the
same magnitudes as their ordinary counterparts.

\end{abstract}


During the past several years it has been recognized that $K$ and $B$
physics provide an exciting test ground for new physics at the TeV region.
Recent reviews may be found in Ref.~\cite{pdg}. In a series of
papers we have studied the possibility that very small mirror matter
admixtures may be present in non-leptonic, weak radiative, and rare mode
decays of strangeness carrying baryons and mesons~\cite{apriori}. Very
recently we showed~\cite{kl2g} that such admixtures may be relevant in
$K_{\rm L}\to\gamma\gamma$ decay. The pattern of these contributions in
the latter decay is very close to the pattern observed in our also recent
analysis of $\Omega^-$ two body non-leptonic decays~\cite{omega2}.
Speacifically, two important properties prevail, the angles $\sigma$,
$\delta$, and $\delta'$ of the mixings of  mirror matter with
ordinary one show a universality property and both strong $SU(3)$ and
$SU(2)$ are obeyed with quite small symmetry breaking corrections in
$\Omega^-$ decays. In $K_{\rm L}\to\gamma\gamma$ only the universality of
$\sigma$ can be tested. It is the purpose of this paper to extend our
previous work on such mixings to $K_{\rm S}\to\gamma\gamma$. In this decay
the universality of $\delta$ and $\delta'$ can be tested, along with the
size of the breaking of the strong symmetries.

Let us first review the current situation on $K_{\rm S}\to\gamma\gamma$.
Its branching ratio has been gradually measured reaching very recently
quite an substantial precision~\cite{lai}. Currently, ${\rm Br}(K_{\rm
S}\to\gamma\gamma)=(2.78\pm0.072)\times10^{-6}$. The theoretical detailed
prediction is at ${\rm Br}(K_{\rm S}\to\gamma\gamma)
=2.1\times10^{-6}$~\cite{lai}. There is an observed enhancement
phenomenon with respect to this  prediction, which covers $\cal{O}$$(p^4)$
of ChPT. After this last experimental  result appeared, an educated guess
about the size of $\cal{O}$$(p^6)$ of ChPT contributions has been made (see
Ref.~\cite{pdg} and reviews cited therein). It is possible that
detailed calculations (which are not yet available) covering such
contributions may either accommodate the new measurement or increase the
difference. For the time being one can only conclude that at present there
is ample room for substantial new physics contributions in this decay. In
case that in the future, it is  demostrated that indeed standard model
contributions saturate the $K_{\rm S}\to\gamma\gamma$ branching ratio,
then one would not be in a position to eliminate mirror matter admixtures,
but one would certainly obtain strong lower bounds for their contributions
in low energy physics (see the closing paragraphs for a more quantitative
discussion).

Let us now proceed to use this phenomenological model in $K_{\rm
S}\to\gamma\gamma$. It is based on parity and flavour admixtures of mirror
matter in ordinary mesons~\cite{apriori}, where the $K_{\rm
S}\to\gamma\gamma$ amplitude is assumed to be enhanced by parity and
flavour conserving amplitudes arising with such admixtures via the
ordinary electromagnetic interaction Hamiltonian as the transition
operator. As a working hypothesis we shall assume that the experimental
branching ratio is saturated by the contribution of such admixtures, and
we shall neglect the Standard Model contributions. This is the same
assumption we have used in our previous work. Of course, this is an
extreme assumption. However, the reason for adopting it is that it
provides a stringent test on such admixtures. Clearly, if a poor or even a
wrong prediction is obtained, then severe constraints on those admixtures
are imposed. Before closing this paper we shall discuss further this
point.

In a model with mirror matter mixings, the physical mesons $K^0_{\rm ph}$
and $\bar{K}^0_{\rm ph}$ with parity and $SU(3)$-flavor violating
admixtures are given by~\cite{apriori}
 
\[
K^0_{\rm ph} =
K^0_{\rm p} -
\frac{1}{\sqrt 2} \sigma \pi^0_{\rm p} +
\sqrt{\frac{3}{2}} \sigma \eta_{\rm 8p} +
\sqrt{\frac{2}{3}} \delta \eta_{\rm 8s} -
\frac{1}{\sqrt 3} \delta \eta_{\rm 1s} -
\frac{1}{\sqrt 2} \delta' \pi^0_{\rm s} +
\frac{1}{\sqrt 6} \delta' \eta_{\rm 8s} +
\frac{1}{\sqrt 3} \delta' \eta_{\rm 1s},
\]

\begin{equation}
\bar{K}^0_{\rm ph} =
\bar{K}^0_{\rm p} -
\frac{1}{\sqrt{2}} \sigma \pi^0_{\rm p} +
\sqrt{\frac{3}{2}} \sigma \eta_{\rm 8p} -
\sqrt{\frac{2}{3}} \delta \eta_{\rm 8s} +
\frac{1}{\sqrt{3}} \delta \eta_{\rm 1s} +
\frac{1}{\sqrt{2}} \delta' \pi^0_{\rm s} -
\frac{1}{\sqrt{6}} \delta' \eta_{\rm 8s} -
\frac{1}{\sqrt{3}} \delta' \eta_{\rm 1s}.
\label{uno}
\end{equation}

\noindent
We have used the $SU(3)$-phase conventions of Ref.~\cite{deswart}. We
recall that the mixing angles $\sigma$, $\delta$, and $\delta'$, are the
parameters of the model, which have been fitted
previously~\cite{detailed}; see later on. The subindeces ${\rm s}$ and
${\rm p}$ refer to positive and negative parity eigenstates, respectively.
 Notice that the physical mesons satisfy $CPK^0_{\rm ph}=-\bar{K}^0_{\rm
ph}$ and $CP\bar{K}^0_{\rm ph}=-K^0_{\rm ph}$.

We can form the $CP$-eigenstates $K_{\rm 1}$ and $K_{\rm 2}$ as

\begin{equation}
K_{\rm 1_{ph}} = \frac{1}{\sqrt{2}} (K^0_{\rm ph} - \bar{K}^0_{\rm ph})
\qquad\mbox{and}\qquad
K_{\rm 2_{ph}} = \frac{1}{\sqrt{2}} (K^0_{\rm ph} + \bar{K}^0_{\rm ph}),
\label{dos}
\end{equation}

\noindent
the $K_{\rm 1_{ph}}$ ($K_{\rm 2_{ph}}$) is an even (odd) state with respect
to $CP$. Here, we shall not consider $CP$-violation and therefore,
$|K_{\rm S,L}\rangle = |K_{1,2}\rangle$.

Substituting the expressions given in Eqs.~(\ref{uno}), we obtain,

\[
K_{\rm S_{ph}} =
K_{\rm S_p} +
\frac{1}{\sqrt{3}} (2\delta + \delta') \eta_{\rm 8s} -
\delta' \pi^0_{\rm s} -
\sqrt{\frac{2}{3}} (\delta - \delta') \eta_{\rm 1s},
\]

\begin{equation}
K_{\rm L_{ph}} =
K_{\rm L_p} -
\sigma \pi^0_{\rm p} +
\sqrt{3} \sigma \eta_{\rm 8p},
\label{tres}
\end{equation}

\noindent
where the usual definitions
$K_{\rm 1_p} = (K^0_{\rm p} - \bar{K}^0_{\rm p})/\sqrt{2}$
and
$K_{\rm 2_p} = (K^0_{\rm p} + \bar{K}^0_{\rm p})/\sqrt{2}$
were used.

Mirror matter admixtures in the physical mesons can contribute to the
$K_{\rm S,L}\to\gamma\gamma$ amplitudes via the ordinary parity and
flavour-conserving electromagnetic interaction Hamiltonian $H^{\rm em}$.
Using Eqs.~(\ref{tres}), a very simple calculation leads to

\begin{equation}
F_{K_{\rm S}\gamma\gamma} =
\frac{1}{\sqrt{3}}(2\delta+\delta')F_{\eta_{\rm 8s}\gamma\gamma} -
\delta' F_{\pi^0_{\rm s}\gamma\gamma} -
\sqrt{\frac{2}{3}}(\delta-\delta')F_{\eta_{\rm 1s}\gamma\gamma},
\label{cuatro}
\end{equation}

\begin{equation}
F_{K_{\rm L}\gamma\gamma} =
-\sigma F_{\pi^0_{\rm p}\gamma\gamma} +
\sqrt{3}\sigma F_{\eta_{\rm 8p}\gamma\gamma},
\label{cinco}
\end{equation}

\noindent
where
$F_{K_{\rm S}\gamma\gamma}=
\langle\gamma\gamma|H^{\rm em}|K_{\rm S_{ph}}\rangle$,
$F_{K_{\rm L}\gamma\gamma}=
\langle\gamma\gamma|H^{\rm em}|K_{\rm L_{ph}}\rangle$,
$F_{\pi^0_{\rm s}\gamma\gamma}=\langle\gamma\gamma
|H^{\rm em}|\pi^0_{\rm s}\rangle$, etc.

Given that $K_{\rm S}$ and $K_{\rm L}$ are $CP=+1$ and $CP=-1$ pure states,
respectively, and because the two-photon state is a $C=+1$ state, then
$K_{\rm S}\to\gamma\gamma$ must go through a so-called parity-violating
transition while $K_{\rm L}\to\gamma\gamma$ goes through a
parity-conserving transition. In the first case the two-photon final state
is $P=+1$ while in the second one, $P=-1$. However, as we can see from
Eqs.~(\ref{cuatro}) and (\ref{cinco}), in the context of mirror matter
admixtures all the contributions to both amplitudes are flavour and parity
conserving. Notice that the additive terms on the right-hand side
of these equations involve only mirror mesons in $F_{K_{\rm
S}\gamma\gamma}$ and only ordinary mesons in $F_{K_{\rm L}\gamma\gamma}$.

We can see from (\ref{cinco}) that the parity-conserving amplitude
$F_{K_{\rm L}\gamma\gamma}$ vanishes in the strong-flavour
$SU(3)$-symmetry limit ($U$-spin invariance):

\begin{equation}
F_{\eta_8\gamma\gamma}=\frac{1}{\sqrt{3}}F_{\pi^0\gamma\gamma}.
\label{cinco1}
\end{equation}

\noindent
The amplitude $F_{K_{\rm S}\gamma\gamma}$ remains non-zero in this limit.
This is the same result previously obtained as a theorem~\cite{limit}.

Let us now concentrate on $K_{\rm S}\to\gamma\gamma$. The experimental data
we shall use comes from Ref.~\cite{pdg}. This includes the average value of
the ${\rm Br}(K_{\rm S}\to\gamma\gamma)=(2.80\pm0.07)\times10^{-6}$, which
differs slightly from the one of Ref.~\cite{lai}. The corresponding
experimental values of the amplitudes that appear in Eq.~(\ref{cuatro}) are
displayed in Table~\ref{table1}. They are obtained using

\begin{equation}
\Gamma(P^0\to\gamma\gamma) =
\frac{\alpha^2m^3_{P^0}}{64\pi^3}F^2_{P^0\gamma\gamma},
\label{seis}
\end{equation}

\noindent
with the decay amplitudes given by the matrix elements
$F_{P^0\gamma\gamma} = \langle\gamma\gamma|H|P^0\rangle$,
and $H=H^{\rm em}$ as the transition operator. The mixing angles are fixed
at $\sigma=(4.9\pm2.0)\times10^{-6}$, $\delta=(2.2\pm0.9)\times10^{-7}$
and $\delta'=(2.6\pm0.9)\times10^{-7}$~\cite{detailed}. To proceed we shall
introduce the assumption of manifest mirror symmetry, as we did in our
previous work, namely, strong and electromagnetic interactions are shared
by mirror and ordinary matter with equal intensity within this
assumption. The underlying QCD dynamics that forms hadrons and makes them
interact is the same for both types, mirror and ordinary ones, as assumed
in the model of Ref.~\cite{barr}. Following this, we assume then

\begin{equation}
|F_{\pi^0_{\rm s}\gamma\gamma}|=
|F_{\pi^0\gamma\gamma}|,
\quad
|F_{\eta_{\rm s}\gamma\gamma}|=
|F_{\eta\gamma\gamma}|,
\quad {\rm and} \quad
|F_{\eta'_{\rm s}\gamma\gamma}|=
|F_{\eta'\gamma\gamma}|.
\label{nueve}
\end{equation}

Mirror mesons will carry the same $SU(2)$ and $SU(3)$ quantum numbers as
ordinary mesons and they will fall into irreducible representations of
these groups, although different from the ordinary ones. These symmetries
are good approximate symmetries, a relationship as Eq.~(\ref{cinco1})
should be valid in the $SU(3)$ symmetry limit for $F_{\pi^0_{\rm
s}\gamma\gamma}$ and $F_{\eta_{\rm 8s}\gamma\gamma}$. This means then that
these two amplitudes must carry the same relative sign determined by this
equation. It also means that Eq.~(\ref{cinco1}) should be obeyed by the
magnitudes of these two amplitudes within a reasonable breaking of
$SU(3)$. We shall assume in what follows that Eq.~(\ref{cinco1}) is obeyed
within an uncertainty of about $15\%$.

One must also expect that the mirror states  $\eta_{\rm s}$ and $\eta'_{\rm
s}$ that appear in Eq.~(\ref{nueve}) follow a mixing scheme between
the states $\eta_{\rm 8s}$ and $\eta_{\rm 1s}$ analogous to the ordinary
$\eta$-$\eta'$ mixing. In the latter case, it has been established that in
general two mixing angles are necessary. One cannot assume that the same
rotation applies to the octet-singlet states and to their decay constants.
For a review see Ref.~\cite{feldmann}. However, we shall use this mixing
only at the amplitude level and in this case only one mixing angle
appears~\cite{cao}. In this respect, it should be clear that we are
not making the questionable assumption that only one mixing angle is used
both for the states and the decay constants. Then, following Ref.~\cite{cao}, we
introduce the rotation

\begin{equation}
\left(
\begin{array}{c}
\eta_{\rm s} \\ \eta'_{\rm s}
\end{array}
\right)
=
\left(
\begin{array}{cc}
\cos{\theta_{\rm s}} & \sin{\theta_{\rm s}} \\
-\sin{\theta_{\rm s}} & \cos{\theta_{\rm s}}
\end{array}
\right)
\left(
\begin{array}{c}
\eta_{\rm 8s}\\
\eta_{\rm 1s}
\end{array}
\right),
\label{siete}
\end{equation}

\noindent
and this leads at amplitude level to

\begin{equation}
\left(
\begin{array}{c}
F_{\eta_{\rm s}\gamma\gamma} \\ F_{\eta'_{\rm s}\gamma\gamma}
\end{array}
\right)
=
\left(
\begin{array}{cc}
\cos{\theta_{\rm s}} & \sin{\theta_{\rm s}} \\
-\sin{\theta_{\rm s}} & \cos{\theta_{\rm s}}
\end{array}
\right)
\left(
\begin{array}{c}
F_{\eta_{\rm 8s}\gamma\gamma} \\ F_{\eta_{\rm 1s}\gamma\gamma}
\end{array}
\right).
\label{ocho}
\end{equation}

We are unable to obtain theoretical predictions for the new amplitudes
involved in Eqs.~(\ref{cuatro}) and~(\ref{ocho}), starting from first
principles. However, our manifest mirror symmetry assumption of
Eq.~(\ref{nueve}) allows us to determine them experimentally, up to a
phase, using the data of Table~\ref{table1}. Finally, for this approach to remain
meaningful it is essential to use the same values of $\delta$ and $\delta'$
determined previously. Otherwise its formulation at a quark level would be
fouled. This latter requires that the mirror matter mixing angles have a
universality property.

From the above formulation of the problem, we are now in a position to make
a prediction for the $K_{\rm S}\to\gamma\gamma$ branching ratio. The
way to proceed is to form a $\chi^2$ function and fit it. This $\chi^2$
contains seven constraints, the four experimental amplitudes of
Table~\ref{table1}, the above mentioned ranges for $\delta$ and $\delta'$, and the
allowed range of $SU(3)$ breaking. Then the six quantities $F_{\eta_{\rm
8s}\gamma\gamma}$, $F_{\pi^0_{\rm s}\gamma\gamma}$, $F_{\eta_{\rm
1s}\gamma\gamma}$, $\delta$, $\delta'$, and $\theta_{\rm s}$ are allowed to
minimized this $\chi^2$. All the two possible signs in front of each
$F_{P^0\gamma\gamma}$ ($P^0=K_{\rm S}, \pi^0, \eta, \eta'$) amplitude must
be explored.

Our best result yields the predictions displayed in Table~\ref{table1}. The
total $\chi^2/({\rm dof})$ is $0.41/1$. The six parameters take the values,
$F_{\eta_{\rm 8s}\gamma\gamma}=(0.617\pm 0.096)\times 10^{-2}\,{\rm
MeV}^{-1}$, $F_{\eta_{\rm 1s}\gamma\gamma}= (-1.610\pm 0.072)\times
10^{-2}\,{\rm MeV}^{-1}$, $\theta_{\rm s}=(17.6\pm 3.7)^{\circ}$,
$F_{\pi^0_{\rm s}\gamma\gamma}=(1.083\pm 0.038)\times 10^{-2}\,{\rm
MeV}^{-1}$, $\delta=(1.80\pm 0.65)\times 10^{-7}$, and $\delta'=(3.01\pm
0.64)\times 10^{-7}$. The $\Delta\chi^2$ contribution of $\delta$ and
$\delta'$ is $0.40$ ($0.20$ each one).

Several remarks are in order. The first one is that the enhanced $K_{\rm
S}\to\gamma\gamma$ amplitude is very well reproduced, along with the other
measured amplitudes, as can be appreciated in Table~\ref{table1}. The
second one is that using the above numbers the ratio $F_{\eta_{\rm
8s}\gamma\gamma}/((1/\sqrt{3})F_{\pi^0_{\rm s}\gamma\gamma})$ of
the amplitudes in the symmetry limit Eq.~(\ref{cinco1}) differs from
one by only $-2\%$. This is indeed very small symmetry breaking. The third
one is that the ranges obtained for $\delta$ and $\delta'$ in this $K_{\rm
S}\to\gamma\gamma$ case overlap very neatly with the ones of our previous work,
mentioned earlier. One can draw the conclusion that manifest mirror
symmetry gives a successful description for the new $K_{\rm
S}\to\gamma\gamma$ decay rate.

Let us now qualify this conclusion carefully. We are certainly not in an
"if and only if" situation. So, let us discuss further our working
assumption that mirror matter admixtures saturate the experimental value
of the  branching ratio. As we have seen such admixtures give a successful
prediction. However, this means only that they are consistent with current
data and in no way can we conclude that they are exclusive. With respect to
this latter point one can envisage three scenarios, namely, (i) only
conventional physics is relevant, (ii) only new physics is relevant, or
(iii) both types of contributions are relevant and should be considered
simultaneously. At present, one cannot reasonably expect (ii) to be true.
However, which of (i) or (iii) is the definitive position is still
premature to tell. The overshooting of the latest experimental
result~\cite{lai} clearly indicates that (i) should be seriously
questioned. The Standard Model predictions are subject to important
uncertainties (see Ref.~\cite{pdg} and references therein) and they
must be put under firm control, before one can establish how much of the
$K_{\rm S}\to\gamma\gamma$ branching ratio they explain and how much they
do not. With respect to (iii), in what concerns mirror matter admixtures,
one should recall the range of $(10^6, 10^7)\ {\rm GeV}$ for the mass of
manifest mirror matter of Ref.~\cite{bound}. Below this range such
admixtures would give unacceptable contributions in low energy physics.
Above it they would become practically irrelevant. For example, if in the
future it is established that Standard Model predictions cover only $2/3$
of the experimental branching ratio of $K_{\rm S}\to\gamma\gamma$, then
the remaining $1/3$ of it would require that $\delta$ and $\delta'$ be
reduced to $60\%$ of the right above values, namely, to $\delta=1.1\times10^{-7}$
and $\delta'=1.8\times10^{-7}$. These values would correspond to changes of
one and one and half standard deviations of the values previously
determined for them, respectively. That is, the values of $\delta$ and
$\delta'$ would still be reasonably compatible with their values from
other decays. In this situation the contributions of mirror matter
admixtures could still be relevant and should not be ignored. If in
contrast conventional contributions do saturate experiment, that is (i)
above prevails, then strong lower bounds on mirror matter would  be set.
Specifically, its mass would not be expected to be below $10^7\ {\rm
GeV}$.

Before closing, this last discussion indicates that more general remarks
should be made. Many new forms of new physics should be considered in low
energies. However, short of discovering what forms of it are out there and
establishing their detailed properties, one may demand that the types of
new physics studied give systematic and consistent effects in low energy
processes. Those types that do this should be taken seriously as
candidates for future discoveries. This far, manifest mirror symmetry
seems to comply with this requirement. Our above analysis complements our
analysis of $K_{\rm L}\to\gamma\gamma$~\cite{kl2g} and the universality of
$\delta$ and $\delta'$ is now explicitly covered, along with the
universality of $\sigma$.

\begin{table}
\caption{Experimentally observed, predicted values and $\Delta\chi^2$
contributions of the $2\gamma$ decay amplitudes of $K_{\rm S}$, $\pi^0$,
$\eta$, and $\eta'$ in ${\rm MeV}^{-1}$. Only the magnitudes of the
experimental values are displayed, the signs for the predictions correspond
to the ones obtained in our best fit.
}
\label{table1}
\begin{tabular}{l c c c}
Decay & Experiment & Prediction & $\Delta\chi^2$ \\
\hline
$K_{\rm S}\to\gamma\gamma$ & $(2.494\pm 0.031)\times10^{-9}$ &
$-2.4938\times10^{-9}$ & $5.9\times10^{-5}$ \\
$\pi^0\to\gamma\gamma$ & $(1.083\pm 0.038)\times10^{-2}$ &
$1.0833\times10^{-2}$ & $5.1\times10^{-5}$ \\
$\eta\to\gamma\gamma$ & $(1.074\pm 0.030)\times10^{-2}$ &
$1.0742\times10^{-2}$ & $2.4\times10^{-5}$ \\
$\eta'\to\gamma\gamma$ & $(1.347\pm 0.069)\times10^{-2}$ &
$-1.3480\times10^{-2}$ & $2.0\times10^{-4}$ \\
\end{tabular}
\end{table}

\end{document}